\documentclass[
    ,final            
  ]
  {aipproc}

\layoutstyle{8x11double}

\begin{document}

\title{Possible Optical/Infrared Jet Emission in 4U 1543-37}

\author{Michelle Buxton}{
  address={Department of Astronomy, Yale University, New Haven CT 06511, U.S.A.}
}

\author{Charles Bailyn}{
  address={Department of Astronomy, Yale University, New Haven CT 06511, U.S.A.}
}

\begin{abstract}
We have taken optical and infrared observations during the 2002 outburst of the soft X-ray transient, 4U 1543-47.  A secondary maximum occurs in the lightcurves during the outburst decline.  This feature is much stronger at infrared wavelengths than optical.  We have applied single blackbody, multicolor blackbody and broken-power law models to the optical/infrared spectral energy distribution of the secondary maximum and find that the broken power-law provides the best fit.  We therefore conclude that the secondary maximum emission originates from a jet.  We also show the most recent lightcurves of the 2002/2003 outburst of GX 339-4 in which a secondary maximum appears.  This leads us to the conclusion that secondary maxima may be a common occurrence in soft X-ray transients during outburst decline which appear after the object transitions into the low-hard state.  Infrared observations of such phenomena will give reliable triggers for multiwavelength observations, allowing us to greatly improve our knowledge of jet formation and behavior, and how this relates to the accretion geometry.
\end{abstract}

\maketitle


\section{Introduction}

The soft X-ray transient (SXT) 4U1543-47 was discovered on 17 August 1971 by the \textit{Uhuru} satellite \cite{mat72}.  Since then, additional X-ray outbursts have been observed in 1983, 1992 and 2002.  The optical counterpart was found during the 1983 outburst by \citet{ped83}.  Dynamical studies of the companion star \cite{oro98b} show that the compact object is most likely a black hole with a mass of 2.7 $\le$ M$_1 \le$ 7.5 M$_{\odot}$. 

The 2002 outburst began on MJD 52442 (UT June 16.683) and was observed by RXTE \cite{mil02,par03}.  During this outburst the soft X-ray light curve exhibited a classical fast rise, exponential decay profile as observed in many other soft X-ray transients \cite{che97}.  The top panel in Figure 1 shows the RXTE ASM light curve from MJD 52390 to 52500.  From early on in the outburst 4U1543-47 was in a thermal-dominated \cite{mcc03} state (also known as high-soft state).  Between MJD 52457-52459 the object briefly transitioned into a low-hard state.  After this time, 4U 1543-47 went back into a thermal-dominated state until MJD 52476 when the source transitioned into the low-hard state \cite{kal02} and remained so until it reached quiescence.  

We observed the 2002 outburst of 4U 1543-47 as part of the YALO\footnote{Now operating as Small and Medium Aperture Research Telescope System (SMARTS), www.astro.yal.edu/smarts/} (Yale, AURA, Lisbon, Ohio) consortium.  Our observations cover most of the outburst cycle in IR and some in optical.  A ``secondary maximum'' appeared in the optical and IR (OIR) light curves which is not apparent in the ASM lightcurve.  In this paper we present our OIR lightcurves and focus our analysis on the secondary maximum.  We argue that this secondary maximum is synchrotron emission which originates from a jet and that this may be a common phenomenon of SXTs during outburst decline.  

\section{Observations and Data Reduction}

$V$- and $J$-band images of 4U1543-47 were obtained on a daily basis (weather permitting) from MJD 52323.3 (UT 2002 February 17) using the YALO 1.0m telescope with ANDICAM\footnote{http://www.astronomy.ohio-state.edu/ANDICAM/}, a dual-channel imager capable of obtaining OIR data simultaneously.  Data was recorded by a Lick/Loral-3 2048x2048 CCD on the optical channel and a Rockwell 1024x1024 HgCdTe "Hawaii" Array on the infrared channel.  On MJD 52431 (UT 2002 June 5) the YALO optical CCD failed.  However, $J$-band observations continued on a daily basis.  When the $J$-band light curve was observed to have risen by $\sim$ 0.5 mag ($\sim$ MJD 52440) we initiated additional observations in the $K$-band.  Optical observations were obtained on a daily basis, when possible, from MJD 52442.9 - 52500.8 (UT 2002 June 18 - August 15) using the 74 inch telescope at Mount Stromlo Observatory with the Cassegrain imager,

a 2K x 4K CCD and $B, V$ and $I$ filters.  Infrared observations ceased when the object was no longer high enough in the night sky to observe.  

OIR data reduction and photometry were performed using the usual routines in IRAF.  Calibration of the optical data was derived by calculating the offsets from a secondary standard, the magnitude of which was provided by Jerry Orosz (private communication).  Infrared images were calibrated via observations of the primary standard P9187\footnote{See Persson faint IR standards at\\www.ctio.noao.edu/instruments/ir\_instruments/ir\_standards/hst.html}, observed on MJD 52520 (UT 2002 September 2).  The optical magnitudes were corrected for airmass extinction using coefficients taken from the CTIO extinction table in IRAF while the infrared atmospheric extinction coefficients were taken from \citet{fro98}.  

\begin{figure}
  \includegraphics[height=.5\textheight]{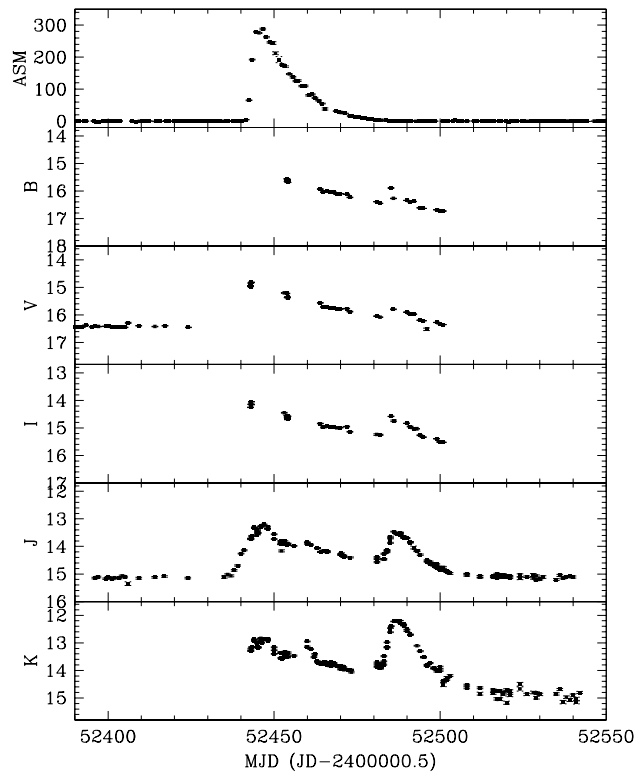}
  \caption{YALO optical and infrared observations of the 2002 outburst of 4U 1543-47.  Top panel is the one-day averaged RXTE ASM lightcurve.  }
\end{figure}

\section{Lightcurve Morphology}

Figure 1 shows the OIR light curves together with the RXTE ASM light curve.  The $J$-band data rises $\sim$ 3-4 days before the ASM, however, this difference will most likely decrease if the analysis is done with the more sensitive PCA data instead.  It is not clear that there is any noticeable difference in the peak times between the ASM and $J$-band.  

After the primary X-ray maximum, the X-rays, OIR lightcurves proceed on a steady decline.  At MJD $\sim$ 52460 the $K$-band lightcurve, and to a lesser extend the $J$-band, experiences a small reflare.  It is interesting to note that just before this reflare 4U 1543-47 briefly transitioned into a low state and low-frequency QPOs were observed \cite{par03}.  The object went back into a thermal-dominated state at MJD 52463.  At MJD 52476, 4U 1543-47 transitioned again into the low state when, again, low-frequency QPOs were observed \cite{par03}.  No optical or infrared data was obtained during the time of transition, however, soon after a third reflare is observed peaking at MJD 52487.  This third reflare, which we call a secondary maximum, is much stronger in the infrared than in the optical and the strength of the peak increases with wavelength.  The reflare is so strong in the $K$-band that it's peak surpasses that of the primary maximum.  Interestingly, no response is seen in the ASM lightcurve.  This is not to say that there is no response in the X-rays at all.  Indeed, preliminary analysis of the PCA data shows a large decrease in the power-law index, QPO frequency and disk blackbody flux occurring close to the time the secondary maximum begins to rise \cite{kal03}.  A radio detection was made at the peak of the secondary maximum at a flux density of 3.8 $\pm$ 0.6 mJy at 843 MHz (Jeff McClintock, private communication).  

After the secondary maximum peak, all lightcurves decline.  The $J$- and $K$-bands eventually follow the decrease from the primary peak until reaching quiescent levels.  
   
\section{Spectral Energy Distribution}

The most extraordinary feature of these lightcurves is the secondary maximum at MJD 52487 and the strength of this feature in $J$ and $K$ compared to the optical bands.  This secondary maximum flux is in addition to the flux from the primary maximum decline since the infrared lightcurves return and follow the primary outburst decline after the reflare.  To ascertain the origin of the secondary maximum we fitted a single blackbody, multicolor blackbody \cite{mit84} and broken power law to the spectral energy distribution of the secondary maximum.  The underlying flux from the primary outburst was removed by fitting a straight line to the data before and after the secondary maximum and subtracting the resulting fit.  Hence, the SED shown in Figure 2 is derived solely from the secondary maximum flux.  

In Figure 2 we also show our best fits of each model to the SED.  The long-dashed line is a single blackbody model of T=1600K.  This model can account for the infrared data but none of the optical.  At this temperature, the emitting region would have a radius of $\sim$ 2 x 10$^{17}$cm.  For an orbital period of 1.123 days \cite{oro98b}, the primary Roche lobe is $\sim$ 5 x 10$^{11}$cm.  Hence, if the secondary maximum emission is thermal, the source would be at a distance 10$^6$ times further than the primary Roche lobe, which seems implausible.

\begin{figure}
  \includegraphics[height=.3\textheight]{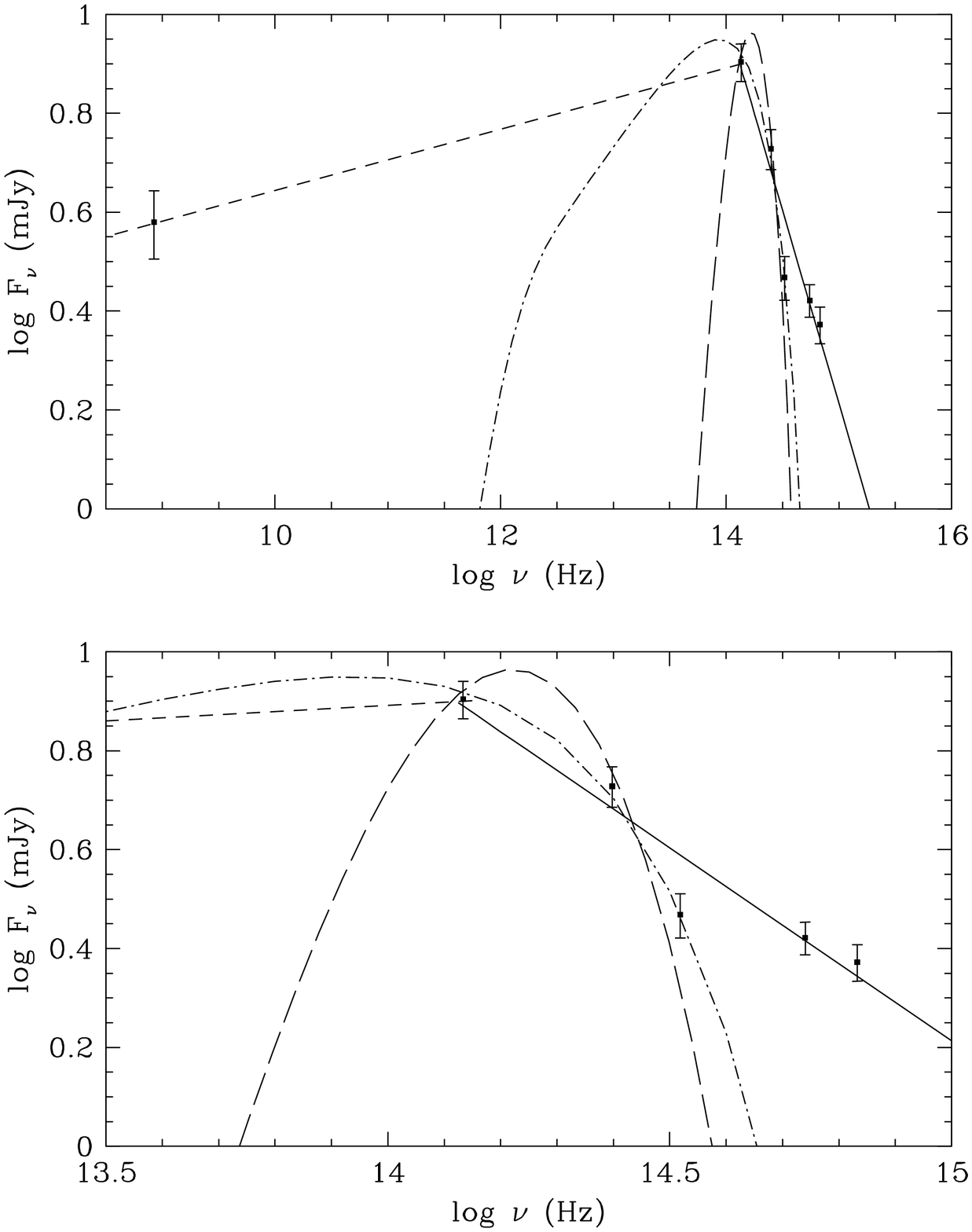}
  \caption{Blackbody and broken power-law fits to the secondary maximum SED.  Top panel shows entire SED, bottom panel shows only OIR data.  Long-dashed line is for single blackbody of T = 1600K.  Dot-dashed line is multicolor blackbody fit (see parameter values in text).  Solid line is best fit power-law to OIR data with $\alpha$ = -0.78 $\pm$ 0.14.  Short-dashed line shows power-law with $\alpha$ = 0.062 $\pm$ 0.003, taken to be a lower limit.}
\end{figure}

The dot-dashed line in Figure 2 represents our best multicolor-blackbody fit.  The fit parameters are as follows:  $r_{out}$ = 10$^{5.65}r_{in}$ (where $r_{in}$ was fixed to 3 gravitational radii), $T_{out} \sim 20^o$K and M\_dot = 10$^{12}$ gs$^{-1}$ (see Mitsuda et al. 1984 for parameter definitions).  We fixed the mass of the black hole at 5 M$_{\odot}$ \cite{oro98b}.  As for the single blackbody case, the model can account for some but not all of the optical/infrared data and it is clear that the outer temperature given by the best fit is too cold for a physical disk.

The broken power-law fit ($F_\nu \propto \nu^{\alpha}$) is shown as a dashed-solid line in Figure 2.  We use this broken power-law to represent emission from a synchrotron source.  The solid line is a linear least-squares fit (in log space) to the OIR data, giving $\alpha$ = -0.78 $\pm$ 0.14.  Since we have only one radio datapoint, and since the SED must break somewhere between the radio and $K$-band points, we can only give a lower-limit of $\alpha$ $\ge$ 0.062 $\pm$ 0.003 to the first part of the SED.  This is shown as the dashed line in Figure 2.  The broken power-law fit is much improved over the blackbody fits.  

Since the OIR data are best described by the power-law fit, together with the unphysical parameters given by the blackbody fits, we conclude that the most likely source of the reflare is jet synchrotron emission.  This conclusion is supported by the radio detection made at the peak of the secondary maximum.

\section{Other Examples}

This is not the first time that this type of behavior has been observed during the outbust decline of soft X-ray transients.  \citet{jai01a} observed a large reflare in the OIR during the 2000 outburst decline of XTE J1550-564.  Recent observations made by us \cite{bux03a,bux03b} of the 2002/2003 outburst of GX 339-4 (see Figure 3) show the same phenomenon.  In both cases, it appears after the sources transition into the low state and the reflare is much stronger in the infrared than optical.  Past studies of SXT outbursts report a ``plateau'' in optical data during outburst decline \cite[see][for example]{che97}.  These plateau may be the same phenomenon as we see here.  

\begin{figure}
  \includegraphics[height=.4\textheight]{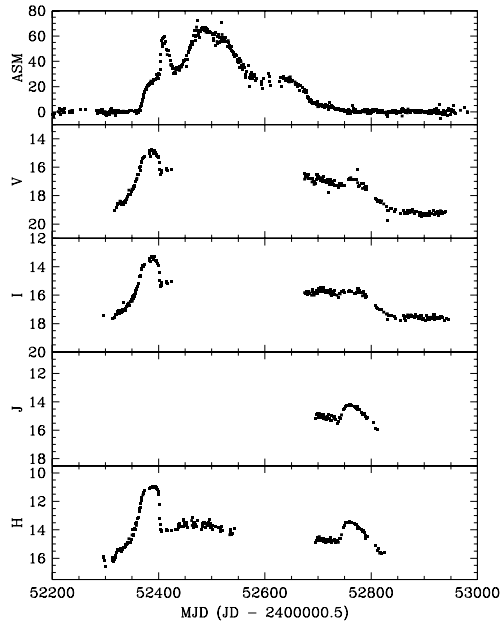}
  \caption{YALO/SMARTS optical and infrared data of GX 339-4 during the 2002/2003 outburst.  Top panel is the one-day averaged RXTE/ASM lightcurve.  During the decline a secondary maximum appears with characteristics similar to those seen in 4U 1543-47 and XTE J1550-564 \cite{jai01a}.}
\end{figure}

It is now widely accepted that jets are a near-ubiquitous property of SXTs during the low state \cite{fen01}.  This scenario is now increasingly supported by OIR observations such as those presented in this paper.  Indeed, since the infrared data show such strong reflares, they may be used as a reliable trigger for other multiwavelength observations.  In doing so we will be able to better constrain when jets appear and make progress in our understanding of jet formation and how this relates to the accretion geometry.

\begin{theacknowledgments}
The authors would like to thank Alfred Chen for assisting in the optical data reduction and Emrah Kalemci and Nick Morgan for their careful perusal of the manuscript.  MB is supported by NSF grant AST-0098421.
\end{theacknowledgments}


\bibliographystyle{aipproc}   



\end{document}